\documentclass[12pt]{amsart}
\usepackage{amsmath,amssymb}

\newtheorem{thm}{Theorem}
\newtheorem{lem}[thm]{Lemma}
\newtheorem{cor}[thm]{Corollary}
\newtheorem{prop}[thm]{Proposition}

\newcommand{\TT}{\mathbf T^2}
\newcommand{\OPN}{\operatorname{Op}_N}
\newcommand{\UN}{U_N}
\newcommand{\mat}{\operatorname{Mat}}
\newcommand{\Up}{U_p}
\newcommand{\Upk}{U_{p^k}}
\newcommand{\TN}{T_N}  
\newcommand{\mnorm}{ {\mathcal N}}
\newcommand{\OF}{{\mathfrak O}}
\newcommand{\OK}{{\mathfrak O}_K}

\newcommand{\Z}{{\mathbf Z}}
\newcommand{\Q}{{\mathbf Q}}
\newcommand{\R}{{\mathbf R}}

\newcommand{\op}{\operatorname{Op}_N}

\newcommand{\norm}[1]{\left\| #1 \right\|_2}
\newcommand{\supnorm}[1]{\left\| #1 \right\|_{\infty}}
\newcommand{\T}{T_{N}}
\newcommand{\HN}{\mathcal H_N}
\newcommand{\Torus}{{\mathcal C}_A}
\newcommand{\Torusev}{\Torus^{\theta}}
\newcommand{\Sol}{\operatorname{Sol}} 
\newcommand{\HH}{{\mathbf H}}
\newcommand{\tr}{\operatorname{tr}}

\begin{document}

\title[Value distribution  for desymmetrized quantum maps]
{Value distribution for eigenfunctions of desymmetrized quantum maps} 

\author{P\"ar Kurlberg and Ze\'ev Rudnick}
\address{
Department of Mathematics, 
University of Georgia, Athens, GA 30602, U.S.A.
({\tt kurlberg@math.uga.edu})}
\address{Raymond and Beverly Sackler School of Mathematical Sciences,
Tel Aviv University, Tel Aviv 69978, Israel 
({\tt rudnick@post.tau.ac.il})}

\date{\today}
\thanks{Partially supported by a grant from THE ISRAEL SCIENCE FOUNDATION
founded by the Israel Academy of Sciences and Humanities, and by 
the National Science Foundation grant no. DMS-0071503}

\begin{abstract}
We study the  value distribution and  extreme values of eigenfunctions  
for the  ``quantized cat map''.  
This is the quantization of a hyperbolic linear map of the torus.   
In a previous paper it was observed that there are 
quantum symmetries of the quantum map - a commutative group of unitary
operators which commute with the map, which we called ``Hecke
operators''. 
The eigenspaces of the quantum map thus admit an orthonormal basis consisting
of eigenfunctions of all the Hecke operators, which we call  ``Hecke
eigenfunctions''. 

In this note we investigate suprema and value distribution of the Hecke
eigenfunctions. For prime values of the inverse Planck constant N 
for which the map is diagonalizable modulo N (the ``split primes'' for
the map), we show that  the Hecke eigenfunctions are 
uniformly bounded and their absolute values
(amplitudes) are either constant or have a semi-circle value
distribution as N tends to infinity. Moreover in the latter case different
eigenfunctions become  statistically independent. 
We obtain these results via the Riemann hypothesis for curves over
a finite field (Weil's theorem) and recent results of N. Katz on
exponential sums.  
For general N we obtain a nontrivial bound on the
supremum norm of these Hecke eigenfunctions.

\end{abstract}

\maketitle


\section{Introduction}

In the past few years, much attention has been devoted to the behavior 
of eigenfunctions of classically chaotic quantum systems. 
One aspect of this topic concerns their value distribution and
specifically their extreme values \cite{Berry, AS, Sarnak-AQC, IS,
Hejhal, ABST}. 
Our aim is to explore this 
for one of the best-studied models in quantum chaotic dynamics, 
the ``quantized cat map'' \cite{HB}.  
This is the quantization of a hyperbolic
linear map $A$ of the torus.    
For a brief background about this model,  see section~\ref{background}. 

For each integer $N\geq 1$ (the inverse Planck constant) 
let $\UN(A)$ denote the quantization of $A$ as
a unitary operator  on $\HN=L^2(\Z/N\Z)$. 
In a previous paper \cite{KR1} it was observed that there are 
{\em quantum symmetries} of $\UN(A)$ - a commutative group of unitary
operators which commute with $\UN(A)$. We called these {\em Hecke
operators}, in analogy with the classical theory of modular forms. 
The eigenspaces of $\UN(A)$ thus admit an orthonormal basis consisting
of eigenfunctions of all the Hecke operators, which we called {\em Hecke
eigenfunctions}. 
These can be thought of as the eigenfunctions for the 
desymmetrized quantum map.  

In \cite{KR1} we showed that the Hecke eigenfunctions become
uniformly distributed as $N\to \infty$. 
In this note we investigate suprema and value distribution of the Hecke
eigenfunctions. For general $N$ we obtain a nontrivial bound on the
supremum norm of these Hecke eigenfunctions. For prime values of $N$
for which $A$ is diagonalizable modulo $N$ (the ``split primes'' for
$A$), we obtain much more refined, optimal results via the modern theory of
exponential sums. We show that for these values of $N$, the Hecke
eigenfunctions  are uniformly bounded and their absolute values
(amplitudes) are either constant or have a semi-circle value
distribution as $N\to \infty$. Moreover in the latter case different
eigenfunctions become  statistically independent. 

Below is a detailed description of our results. 

\subsection{Suprema - general $N$} 
The trivial bound  
\begin{equation}\label{trivial bound}
||\psi||_\infty \leq N^{1/2}
\end{equation}
is a consequence
of the $L^2$-normalization 
$$
||\psi||_2^2 = \frac 1N \sum_{Q\in \Z/N\Z} |\psi(Q)|^2 = 1 \;.
$$

Equidistribution of the eigenfunctions \cite{KR1} implies 
that one can do better: 
$||\psi||_\infty  = o(N^{1/2})$. 
Our first result  gives a quantitative improvement on this: 
\begin{thm}\label{thm 1}
Let $\psi$ be a Hecke eigenfunction,  normalized by 
$||\psi||_2=1$. Then the supremum of $\psi$ satisfies 
$$
||\psi||_\infty \ll_\epsilon N^{3/8+\epsilon}
$$
for all $\epsilon>0$, the implied constant depending only on
$\epsilon$ and not on $\psi$. 
\end{thm}

\subsection{The split primes}
We next consider the case when $N$ is a prime for which 
$A$ is diagonalizable modulo $N$ (these constitute half the primes). 
In this case, the group of Hecke operators is isomorphic to the
multiplicative group 
$(\Z/N\Z)^*$ and the Hecke eigenfunctions  correspond to 
eigencharacters of the group of Hecke operators, namely to 
Dirichlet characters $\chi$ modulo
$N$. For nontrivial $\chi$, the corresponding eigenspace is
one-dimensional, while for the trivial character $\chi_0$ the
eigenspace is two-dimensional. For nontrivial $\chi$ denote by
$\psi_{\chi,N}$ a Hecke eigenfunction of norm 1. 
For simplicity, we also assume that $A$ is not triangular modulo $N$
(which holds for all but finitely many $N$). 
The suprema of the Hecke eigenfunctions in this case are given by 
\begin{thm}\label{thm: uniform boundedness}
If $N$ is a split prime for $A$ such that $A$ is not triangular modulo
$N$ then 
\begin{enumerate}
\item The Hecke eigenspace corresponding to the trivial character has an
orthonormal  basis $\psi_{0,N}, \psi_{\infty,N}$ for which the amplitude is
constant: $|\psi_{\infty,N}| = 1$ and $|\psi_{0,N}| = \sqrt{1-1/N}$. 
\item For nontrivial $\chi$ mod $N$, $||\psi_{\chi,N}||_\infty \leq 
2\sqrt{1-1/N}$. 
\end{enumerate}
\end{thm}

We finally turn to the question of value distribution. The issue here
is only for nontrivial characters $\chi$. We will show that the value
distribution of $\frac 12 |\psi_{\chi,N}|$ tends to the {\em semi-circle
measure}, which is the measure $\mu_{sc}$ on $[0,1]$  such that 
$$
\mu_{sc}(I) = \int_I \frac 4{\pi} \sqrt{1-u^2}du
$$
for any interval $I\subset [0,1]$. This measure is the image of Haar
measure on $SU(2)$ under the map $g\mapsto |\tr(g)|/2$. 

The precise result is:
\begin{thm}\label{thm:value distribution}
If $N$ is restricted to vary only over split primes for $A$ then 
\begin{enumerate}
\item For any nontrivial character
$\chi$ modulo $N$, the amplitude $|\psi_{\chi,N}(t)|$ has a
semi-circle limit distribution as $N\to \infty$, i.e. for any
subinterval $I\subset [0,1]$ we have as  $N\to \infty$ through split
primes for $A$
$$
\frac 1N\#\{t\mod N: |\psi_{\chi,N}(t)|/2\in I \} \to \mu_{sc}(I) \;.
$$
\item For $r\geq 2$, and any choice of  distinct nontrivial characters
$\chi_1$, $\dots$, $\chi_r$, the amplitudes 
$|\psi_{\chi_1,N}|,\dots,|\psi_{\chi_r,N}|$   are statistically
independent, that is for any choice of subintervals $I_1,\dots ,I_r
\subset [0,1]$ we have 
$$
\frac 1N\#\{t\mod N: |\psi_{\chi_i,N}(t)|/2\in I_i, \forall i=1,\dots,r
\} 
\to \prod_{i=1}^r \mu_{sc}(I_I) 
$$
as  $N\to \infty$  through split primes for $A$. 
\end{enumerate}
\end{thm}
Theorem~\ref{thm:value distribution} is a consequence of a 
corresponding theorem by N. Katz \cite{Katz-recent} 
on the value distribution of certain exponential sums, see 
Section~\ref{sec:value dist}. 

\subsection{Comparison with Maass forms} 
To put our results in perspective, we briefly survey the situation 
in another well studied example in quantum chaology - the 
eigenfunctions of the hyperbolic Laplacian $\Delta$ on modular
surfaces. These are the quotient $\HH/\Gamma$ of the hyperbolic plane $\HH$ by 
a congruence subgroup $\Gamma$ of the units of a quaternion algebra (in the
compact case) or of the modular group $SL(2,\Z)$, 
see Sarnak's survey \cite{Sarnak-AQC}. 
In these cases, there is a commuting family of 
self-adjoint operators which commute with the Laplacian, and
thus there is an orthonormal basis $\psi_j$, $ j=0,1,\dots$ 
of $L^2( \HH/\Gamma)$  
consisting of (real-valued) eigenfunctions of the Laplacian 
($\Delta\psi_j+\lambda_j\psi_j = 0$) and of all the Hecke operators. 
These are called Maass-Hecke forms. 
To compare with our result,  note that 
the Laplace eigenvalue $\lambda$ scales with Planck's constant $h$ like
$1/h^{2}$. In the case of the cat map, the inverse Planck constant $N$
equals $1/h$.  

A general bound for the eigenfunctions of the Laplacian on
any compact Riemannian surface  gives  \cite{Ho}: 
$$
||\psi_j||_\infty\ll \lambda_j^{1/4} \sim 1/h^{1/2}
$$
which is analogous to the trivial bound \eqref{trivial bound}. 
Iwaniec and Sarnak \cite{IS} studied the supremum of the Maass-Hecke
forms, and showed that for $L^2$-normalized forms one has 
$$
||\psi_j||_\infty \ll_\epsilon \sqrt{\lambda_j}^{5/12+\epsilon}
\sim 1/h^{5/12+\epsilon}
$$ 
for all $\epsilon>0$. 
This is analogous to our result in Theorem~\ref{thm 1}.  

Unlike in the cat map case,  for  modular surfaces equidistribution
of eigenfunctions is still open, though recent work of Sarnak
\cite{Sarnak-CM} establishes this for a  subsequence of
eigenfunctions of ``CM type'' for congruence subgroups of $SL(2,\Z)$,
and a recent formula of  Watson \cite{Watson} confirms that 
it is implied by the Generalized Riemann
Hypothesis for certain automorphic $L$-functions. 


Concerning the question of value distribution 
for  modular surfaces, 
numerical experiments indicate that the Hecke eigenfunctions have a
locally Gaussian value distribution \cite{Hejhal, AS}.  
In the compact case, this means that if $\psi$ is an $L^2$-normalized
cusp form with  eigenvalue $\lambda>0$, which is an 
eigenfunction of all Hecke operators, then the measure of the set of
$z\in \HH / \Gamma$ where the amplitude $|\psi(z)|<r$ is asymptotic to
$\sqrt{2/\pi}\int_{0}^r e^{-t^2/2} dt$ as $\lambda \to \infty$. 
Here ``measure'' means the hyperbolic measure $dxdy/y^2$ on
$\HH/\Gamma$ normalized to have
total area unity.
Moreover these experiments indicate that eigenfunctions
corresponding to different eigenvalues are statistically
independent\cite{Hejhal, AS}. 

At present, we seem to be far from being able to prove such
statements for modular surfaces (not to mention doing it for a
generic system).  Recently some progress has been made toward Gaussian
value distribution:  Watson \cite{Watson} showed that the
third moment $\int_{\HH/\Gamma} \psi^3$ of the eigenfunctions vanishes
as $\lambda\to \infty$ (all odd moments vanish for the Gaussian
distribution),  
and Sarnak showed that for ``CM forms'', the fourth moment agrees with
the Gaussian moment, i.e. $\int_{\HH/\Gamma}  |\psi|^4 \to 3$ as $\lambda\to
\infty$.  

{\bf Acknowledgments:} We thank P. Sarnak for stimulating
discussions on the subject matter of this paper and N. Katz for 
discussions on exponential sums and for providing us with
\cite{Katz-recent}.

\section{Background}\label{background}

The full details on the cat map and its quantization can be found in
\cite{KR1}. For the reader's convenience  we briefly recall
the setup:

\subsection{Classical dynamics}

The classical dynamics are given by a hyperbolic linear map
$A\in SL(2,\Z)$ so that
$x=(\begin{smallmatrix}p\\q\end{smallmatrix})\in \TT\mapsto Ax$ is a
symplectic map of the torus.  Given an observable $f\in
C^\infty(\TT)$, the classical evolution defined by $A$ is $f\mapsto
f\circ A$, where $(f\circ A)(x)=f(Ax)$. 

\subsection{Kinematics: The space of states} \label{kinematics} 
As the Hilbert space of states, we take distributions $\psi(q)$ on the
line $\R$ which are periodic in both the position and the momentum
representation. This restricts $h$, Planck's
constant, to take only inverse integer values. 
With $h=1/N$, the space of states, denoted $\HN$, is of dimension $N$
and consists of periodic  
point-masses at the coordinates  $q=Q/N$, $Q\in \Z$. 
We identify $\HN$ with $L^2(\Z/N\Z)$, where
the inner product $\langle\,\cdot\,,\,\cdot\,\rangle$ is given by
\begin{equation*}
\langle \phi,\psi \rangle 
 = \frac1N \sum_{Q\bmod N} \phi(Q) \, \overline\psi(Q) .
\end{equation*}
 
\subsection{Observables:} 
The basic observables are given by the operators $\TN(n_1,n_2)$
acting on $\psi \in L^2(\Z/N\Z)$ via: 
\begin{equation}\label{action of T(n)}
\left( \TN(n_1,n_2)\psi \right) (Q) =
e^{\frac {i\pi n_1 n_2}N} e(\frac{n_2Q}N)\psi(Q+n_1).
\end{equation}
For any smooth classical observable 
$f\in C^\infty(\TT)$ with  Fourier expansion 
$$f(x) = \sum_{n_1,n_2\in \Z} \widehat f(n_1,n_2) e(n_1 p + n_2 q),\quad 
x=(\begin{smallmatrix}p\\q\end{smallmatrix})\in \TT,
$$  
its quantization, $\OPN(f)$, is given by
$$
\OPN(f) := \sum_{n_1,n_2\in \Z} \widehat f(n_1,n_2)  \TN(n_1,n_2)
$$

\subsection{Dynamics:}\label{dynamics}

We let $\Gamma(4,2N) \subset SL(2,\Z)$ be the subgroup
of matrices that are congruent to the identity matrix modulo
$4$ (resp., $2$) if $N$ is even (resp., odd).
For $A \in \Gamma(4,2N)$ we can assign unitary operators $\UN(A)$,
acting on $L^2(\Z/N\Z)$, having the following important properties:
\begin{itemize}
\item 
``Exact Egorov'':  For all observables $f \in C^\infty(\TT)$
$$ 
 \UN(A)^{-1}  \OPN(f) \UN(A)= \OPN(f\circ A).
$$
\item The quantization depends only on $A$ modulo $2N$: if $A,B
  \in \Gamma(4,2N)$ and $A  \equiv B \mod 2N$ then  
$$
\UN(A)=\UN(B)
$$
\item The quantization is multiplicative: if $A,B \in
  \Gamma(4,2N)$, then
\begin{equation}\label{muliplicativity}
\UN(AB)=\UN(A)\UN(B)
\end{equation}
\end{itemize}

\subsection{Hecke eigenfunctions}

%

If $\alpha$ is an eigenvalue of $A$, form 
$\OF=\Z[\alpha]$ which is an {\em order} in the real quadratic field
$K=\Q(\alpha)$. (Note that $\OF$ is not necessarily equal to $\OK$,
the full ring of integers in $K$.)
Let $v=(v_1,v_2)\in\OF^2$ be a vector such that 
$vA = \alpha v$. 
Let  $I:=\Z[v_1,v_2] \subset \OF$. Then $I$  is an $\OF$-ideal, and
the matrix of $\alpha$ acting on $I$ by 
multiplication in the basis $v_1,v_2$ is precisely $A$. 
The choice of basis of $I$ gives an identification 
\begin{equation}\label{iota}
\iota:I\to \Z^2 \;.
\end{equation} 
%
The action of $\OF$  on the ideal $I$  by multiplication gives 
a ring homomorphism
$\iota : \OF \to \mat_2(\Z)$ with  the property that the determinant
of $\iota(\beta)$, $\beta\in \OF$, is given by $\mnorm(\beta)$, where 
$\mnorm : \Q(\alpha) \to \Q$ is the norm map. 

Reducing the norm map modulo $2N$ gives a well defined map
$$
\mnorm_{2N} : \OF/2N\OF \to \Z/2N\Z,
$$
and we let
$\Torusev(2N)$ be the elements in the kernel of this map that are
congruent to $1$ modulo $4 \OF$ (resp., $2\OF$) if $N$ is even (resp.,
odd). 

Now, reducing $\iota$ modulo $2N$ gives a  map
$$
\iota_{2N} : \Torusev(2N) \to SL_2(\Z/2N\Z).
$$
Since $\Torusev(2N)$ is commutative, the properties in
section \ref{dynamics} 
imply that 
$$
\{ \UN( \iota_{2N}(\beta) ) : \beta \in \Torusev \}
$$
forms  a family of commuting operators.
Analogously with modular forms, we call these
{\em Hecke operators},
and functions $\psi \in \HN$ that are simultaneous eigenfunctions of
all the Hecke operators are denoted {\em Hecke eigenfunctions}. Note
that a Hecke eigenfunction is an eigenfunction of
$\UN(\iota_{2N}(\alpha))=\UN(A)$.


\section{Proof of Theorem~\ref{thm 1}}

\subsection{Spectral expansions} 
We first display the intensity $|\psi(Q)|^2$ for $\psi\in \HN$ 
as an expectation value of an $N$-dependent observable. 
Choose $f \in C_c^{\infty}(\R)$ so that $f(0)=1$,
$\int_{-\infty}^\infty f(x) \, dx =0$,
and $f$ is supported in $(-1/2,1/2)$. The function
$$
G_N(x) = N \sum_{k \in \Z} f(N(x-k))
$$
is periodic, and its Fourier
coefficients are given by 
$$
\widehat{G_N}(m) = \widehat{f}(\frac mN),
$$
where $\widehat{f}(y) = \int_{-\infty}^\infty f(x)e^{-2\pi ixy}dx$ 
is the Fourier transform of $f$ on the line.
For $Q\in \Z/N\Z$, set 
$$
g_{N,Q}(p,q) = G_N(q-\frac {Q}N) , \quad (p,q)\in \TT \;.
$$
We obtain a function on the torus which is independent of the
momentum variable $p$, and  is strongly localized in  the position
variable $q$ around $Q/N$.  

\begin{lem} 
Let $Q \in \Z/N\Z$. Then for all $\psi\in \HN$
$$
|\psi(Q)|^2 = \langle \op(g_{N,Q}) \psi, \psi \rangle 
$$
\end{lem}
\begin{proof} 
Recall that 
$$
\op(g_{N,Q}) = \sum_{m,n \in \Z} \widehat{g_{N,Q}}(m,n) \T(m,n) \;.
$$
Since
$g_{N,Q}$ is independent of $p$, we have $\widehat{g_{N,Q}}(m,n)=0$ unless
$m=0$, in which case we have 
\begin{equation}\label{fourier of g_{N,Q}}
 \widehat{g_{N,Q}}(0,n)=e(-\frac{nQ}N)\widehat G_N(n) = 
e(-\frac{n Q}N) \widehat f(\frac nN)\;.
\end{equation}
Since $\T(0,n) \psi(Q') = e( nQ'/N ) \psi(Q')$, we get 
\begin{equation*}
\begin{split}
 \op(g_{N,Q}) \psi (Q') &= 
\sum_{m,n \in \Z} \widehat{g_{N,Q}}(m,n) \T(m,n) \psi(Q') \\
&= \sum_{n \in \Z} \widehat{g_{N,Q}}(0,n) \T(0,n) \psi (Q') \\
&=\sum_{n \in \Z} \widehat{g_{N,Q}}(0,n) e(nQ'/N)\psi(Q') \\
& = g_{N,Q}(0,Q')\psi(Q')  = G_N(\frac{Q'-Q}N) \psi(Q')
\end{split}
\end{equation*}
Since the support of $G_N(x)$ is contained in $(-1/2N,1/2N)$ modulo 1,
and $G_N(0)=N$, we get  
$$
\op(g_{N,Q}) \psi (Q')  = 
\begin{cases}  
               N\psi(Q)& \text{if } Q'=Q \mod N\\ 
               0&\text{otherwise.} 
\end{cases}
$$
Hence
$$
\langle 
\op(g_{N,Q}) \psi, \psi 
\rangle 
=
\frac{1}{N}
\sum_{Q' \mod N}G_N(\frac{Q'-Q}N)  |\psi(Q')|^2 = |\psi(Q)|^2
$$
\end{proof}

\begin{lem} 
\label{l:cauchy-twice}
Let $\psi_1,\dots ,\psi_N\in \HN$. Then  for all
$Q\in \Z/N\Z$ 
$$
\sum_{j=1}^N |\psi_j(Q)|^8 \leq 
\left( \sum_{n \in \Z} |\widehat f(\frac nN)|
\left( \sum_{j=1}^N | \langle \T(0,n) \psi_j,  \psi_j \rangle |^4 
\right)^{1/4} \right)^4
$$
\end{lem}
\begin{proof} 
For ease of notation we put 
$$
t_j(n) = | \langle \T(0,n) \psi_j,  \psi_j \rangle |.
$$
By the previous lemma, 
\begin{equation*}
\begin{split}
|\psi_j(Q)|^2 &=  \langle \op(g_{N,Q}) \psi_j,  \psi_j \rangle \\
&\leq \sum_{n \in \Z} |\widehat{g_{N,Q}}{(0,n)}| t_j(n) \\
&= \sum_{n \in \Z} |\widehat f(\frac nN) | t_j(n)
\end{split}
\end{equation*}
on using \eqref{fourier of g_{N,Q}}. Thus 
\begin{equation*}
\begin{split}
\sum_{j=1}^N |\psi_j(Q)|^8 &\leq 
\sum_{j=1}^N \left( \sum_{n \in \Z} |\widehat f(\frac nN)| t_j(n)
\right)^4 \\
&= \sum_{j=1}^N \sum_{n_1, n_2, n_3, n_4 \in \Z} 
\prod_{k=1}^4  |\widehat f(\frac {n_k} N)| t_j(n_k) \;.
\end{split}
\end{equation*}
Applying Cauchy-Schwartz twice we get that
\begin{equation*}
\begin{split}
\sum_{j=1}^N  t_j(n_1) t_j(n_2) t_j(n_3) t_j(n_4) 
&\leq
\sqrt{ \sum_{j=1}^N  t_j(n_1)^2 t_j(n_2)^2 }
\sqrt{ \sum_{j=1}^N t_j(n_3) t_j(n_4) } \\
&\leq \prod_{k=1}^4  
\left( \sum_{j=1}^N  t_j(n_k)^4 \right)^{1/4}
\end{split}
\end{equation*}
and hence
\begin{equation*}
\begin{split}
\sum_{j=1}^N |\psi_j(Q)|^8 &\leq 
\sum_{n_1, n_2, n_3, n_4 \in \Z} \prod_{k=1}^4
\left(
|\widehat f(\frac {n_k} N)|
\left( \sum_{j=1}^N  t_j(n_k)^4 \right)^{1/4}
\right) \\
&=
\left( 
\sum_{n \in \Z}|\widehat f(\frac n N)| 
\left( \sum_{j=1}^N  t_j(n)^4 \right)^{1/4}
\right)^4
\end{split}
\end{equation*}
\end{proof}

\subsection{A counting problem}
\label{s:fourth-moment}

Recall that we identified the action of the matrix $A$ on $\Z^2$ with
multiplication by its eigenvalue $\alpha$ on the ideal 
$I\subseteq \Z[\alpha]$. 
Let $\iota:I\to \Z^2$ be the identification given in \eqref{iota}.  
We will need the following Proposition  (\cite[Proposition 11]{KR1}):
\begin{prop} \label{prop reduce to count}
Fix  $\nu\in I$. 
Then for any 
orthonormal basis of Hecke eigenfunctions $\psi_j$,  
$$
\sum_{j=1}^N |\langle \TN(\iota(\nu))\psi_j,\psi_j \rangle |^4 \leq \\
\frac {N}{|\Torusev(2N)|^4}\Sol(N,\nu)
$$
where $\Sol(N,\nu)$ is the number of solutions of the equation
\begin{equation}
  \label{eq:basic-mod-N}
  \nu( \beta_1 -\beta_2 +\beta_3 -\beta_4) \equiv 0 \mod N
\end{equation}
with $\beta_1, \ldots, \beta_4$ in $\Torusev(N)$.
\end{prop}

%
%
It was also shown that
equation~\eqref{eq:basic-mod-N} has less 
than $N^{2+\epsilon}$ solutions for $\nu$ fixed, as $N$
tends to infinity. However, in the current setup, we need to make the
dependence on $\nu$ more explicit\footnote{If $n \equiv 0 \mod  
N$, then there are about $N^4 $ solutions; it is essential to
control the contribution from such terms.}. 
Recall that $\alpha$ is an eigenvalue of $A$, and that $\mnorm :
\Q(\alpha) \to \Q$ is the norm map.

\begin{lem} 
\label{l:tj-moment-bound}
We have
$$
\sum_{j=1}^N |\langle \TN(0,n)\psi_j,\psi_j \rangle |^4 \leq 
\gcd(n,N)^{2} N^{-1+\epsilon}
$$
\end{lem}
\begin{proof} 
It is sufficient to show that equation~\ref{eq:basic-mod-N} has less
than 
$$
\gcd(n,N)^{2} N^{2+\epsilon}
$$
solutions.
We argue as in \cite{KR1}, section 7, except for a small twist: If
$(0,1) \in \Z^2$ corresponds to $\omega \in I$, then $(0,n)$
corresponds to $\nu = n\omega$ since the action is $\Z$-linear.  We
may thus make $\nu$ Galois stable by multiplying by
$\overline{\omega}$. Since $\omega$ is in $\Z[\alpha]$, the number
of solutions to
$$
\nu( \beta_1 -\beta_2 +\beta_3 -\beta_4) \equiv 0 \mod N, 
\ \ \beta_i \in \Torusev(N)
$$
is bounded by the number of solutions of 
$$
n \mnorm(\omega)( \beta_1 -\beta_2 +\beta_3 -\beta_4) \equiv 0 \mod N,
\ \ \beta_i \in \Torusev(N)
$$
which in turn is given by the number of solutions to
\begin{equation}
  \label{eq:solutions}
\beta_1 -\beta_2 +\beta_3 -\beta_4 \equiv 0 
\mod \left(N/\gcd(N,n \mnorm(\omega)) \right),
\ \ \beta_i \in \Torusev(N).
\end{equation}
Let $N' = N/\gcd(N,n \mnorm(\omega))$. The number of
solutions of equation~\ref{eq:solutions} then equals the product of:
\begin{itemize}
\item 
The number of solutions to 
$$
\beta_1' -\beta_2' +\beta_3' -\beta_4' \equiv 0 \mod N'
\ \ \beta'_i \in \Torusev(N').
$$
where $\beta_1', \ldots, \beta_4'$ ranges over all elements in
$\Torusev(N')$. 

\item 
The number of elements 
$(\beta_1, \beta_2, \beta_3, \beta_4)  \in \Torusev(N)^4$ 
that reduce to the same element
$(\beta_1', \beta_2', \beta_3', \beta_4')  \in \Torusev(N')^4$.
\end{itemize}
 From proposition 14 in \cite{KR1}, applied with $\nu=1$, it
follows that the first term is $\ll (N')^{2+\epsilon}.$
For the second term, lemma 20 in \cite{KR1} gives that
the cardinality of the cokernel of the reduction map $\Torusev(N) \to
\Torusev(N')$ is uniformly  bounded in $N$. Hence there are 
$\ll (\Torusev(N)/\Torusev(N'))^{4}$
elements $(\beta_1, \beta_2, \beta_3, \beta_4)$ that reduce to 
$(\beta_1', \beta_2', \beta_3', \beta_4')$ modulo $N'$.

Finally, from lemma 8 in \cite{KR1} we have
$$
N^{1-\epsilon} \ll |\Torusev(N)| \ll N^{1+\epsilon}
$$
and thus 
$$\Torusev(N)/\Torusev(N') \ll (N/N')^{1+\epsilon}.$$
The number of solutions is therefore bounded by
$$
(N')^{2+\epsilon} \times (N/N')^{4+\epsilon} 
\ll
N^{2+\epsilon}   \gcd(n \mnorm(\omega),N)^{2+\epsilon} 
\ll
N^{2+2\epsilon}  \gcd(n ,N)^{2} \;.
$$

\end{proof}

\begin{lem} 
\label{l:fourier-chop}
For all $\epsilon>0$, 
$$
\sum_{n \in \Z} |\widehat f(\frac nN)| \gcd(n,N)^{1/2}
\ll_{\epsilon} N^{1+\epsilon}.
$$
\end{lem}
\begin{proof} 
Let $R=1/\epsilon$. Since $f$ is smooth and compactly supported,  
$|\widehat f(\frac nN)| \ll 1$ for $|n| \leq N^{1+1/R}$, 
and $|\widehat f(\frac nN)| \ll \frac{1}{(n/N)^R}$ for $|n|\geq 
N^{1+1/R}$. 

Trivially, $\gcd(n,N)^{1/2} \leq N$, so 
\begin{equation*}
\begin{split}
\sum_{|n| \geq  N^{1+1/R}} |\widehat f(\frac nN)| \gcd(n,N)^{1/2} 
&\ll\sum_{|n|\geq  N^{1+1/R}}
\frac{N}{(n/N)^R} \\
&=N^{R+1} \sum_{|n| \geq  N^{1+1/R}}n^{-R} \\
&\ll 
\frac{N^{R+1}}{(N^{1+1/R})^{R-1}}
=
N^{1+1/R}
\end{split}
\end{equation*}

For the sum over small $n$, we have 
\begin{equation}\label{reduce to gcd}
\begin{split}
\sum_{|n| \leq N^{1+1/R}}
|\widehat f(\frac nN)| \gcd(n,N)^{1/2} &\ll
\sum_{|n| \leq N^{1+1/R}} \gcd(n,N)^{1/2}  \\
&\ll
N^{1/R} \sum_{n=0}^{N-1} \gcd(n,N)^{1/2} \;.
\end{split}
\end{equation}
We note that 
\begin{equation}\label{l:gcd-average-bound}
\sum_{n=0}^{N-1} 
\gcd(n,N)^{1/2}
\ll N^{1+\epsilon} \;.
\end{equation}
Indeed, 
$$
\sum_{n=0}^{N-1} 
\gcd(n,N)^{1/2}
=
\sum_{d|N}
d^{1/2}
\sum_{n \leq N, \gcd(n,N)=d} 1
\leq
\sum_{d|N} d^{1/2}
\frac{N}{d}
=
N \sum_{d|N} d^{-1/2}.
$$
Now $\sum_{d|N} d^{-1/2}$ is bounded by the number of divisors of $N$,
and hence is $\ll N^\epsilon$ for all $\epsilon>0$. 


Therefore from \eqref{reduce to gcd} and \eqref{l:gcd-average-bound} we get 
$$
\sum_{|n| \leq N^{1+1/R}}
|\widehat f(\frac nN) | \gcd(n,N)^{1/2} \ll
\sum_{|n| \leq N^{1+1/R}} \gcd(n,N)^{1/2}  
\ll
N^{1/R+1+\epsilon} \;.
$$
\end{proof}


\subsection{Conclusion of the proof} 
Let $\{\psi_j\}_{i=1}^{N}$ be an orthonormal basis of $\HN$ such that
$\psi_1=\psi$, and each 
$\psi_j$ is a Hecke eigenfunction. We will then bound 
$|\psi(Q)|^8$ trivially by the sum
$\sum_{j=1}^N |\psi_j(Q)|^8$.

By lemma~\ref{l:cauchy-twice},
$$
\sum_{j=1}^N |\psi_j(Q)|^8 \leq 
\left( \sum_{n \in \Z} |\widehat f(\frac nN)|
\left( \sum_{j=1}^N | \langle \T(0,n) \psi_j,  \psi_j \rangle |^4 
\right)^{1/4} \right)^4
$$
and from Lemma~\ref{l:tj-moment-bound} we have
$$
\sum_{j=1}^N | \langle \T(0,n) \psi_j,  \psi_j \rangle |^4 \leq 
\gcd(n,N)^2  N^{-1+\epsilon}
$$
Lemma~\ref{l:fourier-chop} then gives that
\begin{equation*}
\begin{split}
\sum_{j=1}^N |\psi_j(Q)|^8 
&\ll
N^{-1+\epsilon}
\left( \sum_{n \in \Z} |\widehat f(\frac nN)| \gcd(n,N)^{1/2}
\right)^4 \\
&\ll
N^{-1+\epsilon} \left( N^{1+\epsilon} \right)^4
\ll
N^{3+5\epsilon}
\end{split}
\end{equation*}
and thus
$$
|\psi(Q)| \ll N^{3/8+\epsilon} 
$$
as required. 
\qed

\section{Uniform boundedness for split primes}\label{sec: split}


\subsection{Explicit construction of Hecke eigenfunctions}

Let $N=p$ be an odd ``split'' prime, i.e. a prime $p$ which does not divide
$(\tr A)^2-4$ such that $A$ is
diagonalizable modulo $p$. We also assume that $A$ is not triangular
mod $p$,  that is $p$ does not divide any of the off-diagonal entries
of $A$. 
For such $p$, we  give an explicit construction of the Hecke eigenfunctions 
(see \cite{DEGI} for an alternative approach to constructing these). 
This construction will enable us to prove Theorems~\ref{thm: uniform
boundedness} and~\ref{thm:value distribution}. 

The condition above implies that 
$N=p$ is an  odd prime that splits in $K$ and does not
divide the conductor\footnote{Recall that $\OF$ can be
  written as $\Z + f \OK$; the integer $f$ is called the conductor
  of $\OF$.} of $\OF$. 
Since $p$ does not divide the conductor of $\OF$, we have
$\OF/p\OF \simeq \OK/p\OK$. Moreover, since $p$ splits in
$K$, $(\OK/p\OK)^\times \simeq (\Z/p\Z)^\times \times (\Z/p\Z)^\times$, and
from lemma 19 in \cite{KR1} we get that
$\Torusev(p) \simeq (\Z/p\Z)^\times$. On the other hand, since $p$ is odd,
we have $\Torusev(2p) \simeq  \Torusev(p)$, hence $\Torusev(2p)$
is a cyclic group  of order $p-1$.
Let $\beta$ be the generator of this group, and let
$B \in \Gamma(4,2p)$ be congruent to $\iota(\beta)$ modulo $2p$. 
The Hecke operators are then given by $\UN(B^k)$.
Since the order of $B$ modulo $p$ is $p-1$, $B$  can be diagonalized
modulo $p$ (all elements in $(\Z/p\Z)^\times$ are $p-1$ roots of unity.)
Let $(B_2,B_p)$ denote the
mod $2$ and mod $p$ reductions of  $B$. 
Since $B \in \Gamma(4,2p)$ we find that $B_2$ is the identity
matrix, and as $B_p$ is diagonalizable 
there exists $D,M \in
\Gamma(4,2p)$ such that  
$D \equiv \begin{pmatrix} t & 0 \\ 0 & t^{-1}\end{pmatrix} \mod 2p$
is diagonal and 
$$
B \equiv  M D M^{-1} \mod 2p
$$
Because all matrices lie in $\Gamma(4,2p)$, the
multiplicativity property \eqref{muliplicativity} implies that
$\UN(B)=\UN(M)\UN(D)\UN(M)^{-1}$. 

Now, $\UN$ is constructed as a tensor product $\UN = \otimes_{p^k ||
  N} \Upk$ and since $B_2$ is trivial, we find the action is
  determined by $B_p$. 
Recall from section 4 of \cite{KR1} that the action of diagonal
matrices on $\psi \in L^2(\Z/p\Z)$ is given by  
$$
\left(\Up \left( D \right) \psi \right) (x) =
\Lambda_p(t) \psi(tx)
$$
where $\Lambda_p(t)$ is the quadratic character of $(\Z/p\Z)^\times$. Thus, if
$\chi$ is a character on $(\Z/p\Z)^\times$, extended to $\Z/p\Z$ by letting
$\chi(0)=0$, we find that $\chi$ is an eigenfunction of
$\Up \left( D \right)$. We also find that $\delta(x)$, where
$\delta(0)=1$ and $\delta(x)=0$ for $x \neq 0$, is an eigenfunction. 

If $f$ is an eigenfunction of $\Up(D)$, then $\Up(M)f$ is an
eigenfunction of $\Up(B)$ since
$$
\Up(B) \Up(M)f = 
\Up(M) \Up(D) \Up(M^{-1}) \Up(M)f =
\Up(M) \Up(D) f
$$
But $\Up(B)$ generates the group of Hecke operators, hence any  
Hecke eigenfunction will either be of the form 
$\Up(M) \chi$, for $\chi$ nontrivial, or a linear combination
of $\Up(M) \chi_0$ and $\Up(M) \delta$ for $\chi_0$ trivial. 

\subsection{A reduction to exponential sums}

We first note that  $M$ is not upper triangular modulo $p$, otherwise
the same would hold for $B$. Since $A$ is a power
of $B$ (modulo $p$), this would imply that  $A$ is upper triangular
modulo $p$, which is contrary to our assumption on $p$. 
Thus  we may use the Bruhat decomposition of
$SL(2,\Z/p\Z)$ to write 
\begin{equation}\label{bruhat}
M = 
\begin{bmatrix} 1 & b_1 \\ 0 & 1\end{bmatrix}
\begin{bmatrix} 0 & 1 \\ -1 & 0\end{bmatrix}
\begin{bmatrix} 1 & b_2 \\ 0 & 1\end{bmatrix}
\begin{bmatrix} t & 0 \\ 0 & 1/t\end{bmatrix}
\end{equation}
for some $b_1, b_2, t$ (depending on $p$). 
 From section 4 of \cite{KR1}, we obtain that
for $\psi \in L^2(\Z/p\Z)$, 
\begin{equation} 
\label{e:explicit-eigenfn}
(\Up(M) \psi )(x) =
\Lambda_p(t)
\frac{e_p(r_pb_1 x^2)}{\sqrt{p}}
\sum_{y=1}^p 
e_p(r_p (b_2 y^2 + 2xy) ) \psi(ty)
\end{equation}
where $r_p$ is the inverse of $2\mod p$, and $e_p(x):=e^{2\pi i x/p}$.

\subsubsection{The case $\psi = \Up(M) \chi$}
We begin with the following lemma on exponential sums:
\begin{lem}
\label{l:weil-bound}
If $r_p \not \equiv 0 \mod p$ then 
\begin{enumerate}
\item If $\chi$ is nontrivial and $b_2\neq 0 \mod p$ then 
$$
\left|
\sum_{y=1}^p 
e_p(r_p (b_2 y^2 + 2xy) ) \chi(y)
\right|
\leq 2\sqrt{p} 
$$
\item If $\chi=\chi_0$ is trivial or $b_2 =0 \mod p$ then 
$$
\left| \sum_{y=1}^{p-1} e_p(r_p (b_2 y^2 + 2xy) ) \chi(y) 
\right| =\sqrt{p} 
$$
\end{enumerate}
\end{lem}
\begin{proof}
If $\chi$ is trivial or $b_2 \equiv 0 \mod p$ then we can express the sum 
as a classical Gauss sum, and in this case the result is well known.
For $\chi$ nontrivial, we note that the degree of $r_p (b_2 y^2 +
2xy)$ is coprime to $p$, hence we may apply Weil's bound \cite{Weil} 
on exponential sums
(see \cite[page 45, Theorem 2G ]{Schmidt-book}) to obtain
$$
\left| \sum_{y=1}^p 
e_p(r_p (b_2 y^2 + 2xy) ) \chi(y) \right|
\leq 2 \sqrt{p}
$$
Note that the bound is independent of the order of $\chi$!
\end{proof}

\begin{cor}
\label{l:chi-bound}
Let $\psi = \sqrt{p/(p-1)}\Up(M) \chi$. Then  
$\norm{\psi}=1$, $|\psi| \equiv \sqrt{p/(p-1)}$ if $\chi $ is the
trivial character, and for nontrivial characters $\chi$ 
$$
\supnorm{\psi} \leq 2 \sqrt{p/(p-1)} \;.
$$ 
\end{cor}

\subsubsection{The case $f = \Up(M) \delta$}

\begin{lem}
\label{l:delta-bound}
Let  $\psi_{\infty,p}= \sqrt{p}  \Up(M) \delta$. 
Then $\norm{\psi_{\infty,p}} = 1=\supnorm{\psi_{\infty,p}} $.
\end{lem}
\begin{proof}
$\norm{\psi_{\infty,p}}=1$ since $\norm{\delta}^2 = 1/p$ and $\Up(M)$ is
unitary. From equation \ref{e:explicit-eigenfn} we get 
\begin{multline*}
|\psi_{\infty,p}(x)| = 
\left|
\sqrt{p}
(\Up(M) \delta )(x)
\right| \\
=
\left|
\sqrt{p} 
\Lambda_p(t)
\frac{e_p(r_pb_1 x^2)}{\sqrt{p}}
\sum_{y=1}^p 
e_p(r_p (b_2 y^2 + 2xy) ) \delta(ty) 
\right|
= 1
\end{multline*}
since $\delta(ty) = 0$ unless $y=0$. Hence $\supnorm{\psi_{\infty,p}} = 1$. 
\end{proof}
Theorem~\ref{thm: uniform boundedness} follows immediately from
Corollary~\ref{l:chi-bound}  and Lemma~\ref{l:delta-bound}.

%

\section{Value distribution for split primes} \label{sec:value dist}
Let $p$ be a split prime for our map $A$.  We assume that $A$ is not
triangular mod $p$. 
To prove Theorem~\ref{thm:value distribution}, 
we again use  \eqref{e:explicit-eigenfn}, which says that 
we can write the
normalized Hecke eigenfunctions $\psi_{\chi,p}$ for nontrivial $\chi$ as
$$
\psi_{\chi,p}(x)  = \frac{\Lambda_p(t_p)e_p(r_p b_{1,p} x^2)} {\sqrt{p-1}} 
\sum_{y\mod p} e_p(r_p(b_{2,p}y^2+2xy)) \chi(y)
$$
where $r_p$ is the inverse of $2$ modulo $p$, $\Lambda_p$ is the
unique quadratic character mod $p$, and $t_p$, $b_{1,p}$ and $b_{2,p}$
come from the Bruhat decomposition \eqref{bruhat} of the diagonalizing
matrix $M_p$ for $A$. 

Note that $b_{2,p} \neq 0 \mod p$ for  $p$ as in our assumptions, 
since otherwise from \eqref{bruhat}, we find that 
$$
M_p=\begin{bmatrix} -b_{1,p}t_p& 1/t_p \\ -t_p&0\end{bmatrix}
$$
and consequently the matrix $B$ is upper triangular:
$$
B = MDM^{-1} = \begin{bmatrix}  1/t_p& b_{1,p}(t_p-1/t_p) \\ 0 & t_p 
\end{bmatrix} \;.
$$
Since $A$ is a power of $B \mod
p$, this implies that $A$ is also upper triangular, 
contradicting our assumption on $p$ .

Thus  we may express the absolute value of $\psi_{\chi,p}$ in
terms of the exponential sums 
$$
H_p(\chi,R)(t) = \sum_{y\mod p} e_p(R(y^2+ty) ) \chi(y) \;.
$$
First define (following N. Katz) a normalization $F_p(\chi,R)(t)$ 
of these sums as ($R\neq 0 \mod p$): 
$$
F_p(\chi,R)(t) =  - \frac{e_p(Rt^2/8)}
{\sqrt{\chi(-1/2)G(R,\chi)G(R,\Lambda_p)}} H_p(\chi,R)(t)
$$
where $G(R,\chi) = \sum_{x\mod p} \chi(x)e_p(Rx)$ are Gauss sums, and
$\sqrt{\ast}$ denotes any choice of the square root. 
Then we have: 
\begin{equation}\label{relation}
|\psi_{\chi,p}(x) |  =  \frac 1{\sqrt{1-1/p}}
\left | F_p(\chi,r_p b_{2,p})(\frac{2x}{b_{2,p}} ) \right|  \;.
\end{equation}

Concerning the  the normalized sums $F_p(\chi,R)(t)$, 
Katz proved  the following 
value distribution and statistical independence theorem: 
\begin{thm}[\cite{Katz-recent} ]
Let $p$ be an odd prime, 
$\chi$ be a nontrivial character mod $p$ and $R \neq 0 \mod p$. Then
\begin{enumerate}
\item The normalized sums  $F_p(\chi,R)(t)$ are real and take values in
the interval $[-2,2]$. 
\item As $p\to \infty$, the $p$ numbers $\{F_p(\chi,R)(t)/2: t\mod p\}$
become equidistributed in $[-1,1]$ with respect to the semi-circle
measure $\frac 2\pi \sqrt{1-u^2}du$. 
\item For any $r\geq 2$, and a choice of $r$ distinct nontrivial characters
$\chi_1,\dots ,\chi_r$, the $p$ vectors 
$$
\{(F_p(\chi_1,R)(t)/2, \dots ,F_p(\chi_r,R)(t)/2 ): t\mod p\}
$$ 
become equidistributed in $[-1,1]^r$ with respect to the product of the
semi-circle measures. 
\end{enumerate}
\end{thm}

By virtue of the relation \eqref{relation} between the normalized
eigenfunctions $f_{\chi}$ and the normalized sums $F_p(\chi,R)$,
Theorem~\ref{thm:value distribution} is an immediate consequence of
Katz's theorem and the following general lemma:
\begin{lem}
Let $\{f_p(t), t=1,\dots ,p\}$ be a sequence of $p$  points in the 
interval $[0,1]$ which 
become equidistributed as $p\to\infty$ with respect to a probability measure
$\rho(x)dx$ having a continuous density $\rho$. Suppose
$\{g_p(t):t=1,\dots, p \}$ is another sequence of points so that 
$g_p(t) = \theta_p f_p(t)$, with $\theta_p = 1+o(1)$. Then 
$\{g_p(t):t=1,\dots, p \}$ is also equidistributed in $[0,1]$ with
respect to $\rho(x)dx$.
\end{lem}
We leave the proof of this as a simple exercise for the reader.


\begin{thebibliography}{99}


\bibitem[ABST]{ABST}
R. Aurich, A. Backer, R. Schubert and M. Taglieber 
{\em Maximum norms of chaotic quantum eigenstates and random
waves}, Phys. D {\bf 129} (1999), no. 1-2, 1--14. 

\bibitem[AS]{AS}
R. Aurich and F. Steiner {\em Statistical properties of highly excited quantum
eigenstates of a strongly chaotic system}, 
Phys. D {\bf 64} (1993), no. 1-3, 185--214.

\bibitem[Be]{Berry}
M.V. Berry {\em Regular and irregular semiclassical wavefunctions}, 
J. Phys. A {\bf 10} (1977), pp. 2083--2091. 

\bibitem[DEGI]{DEGI} M. Degli Esposti, S. Graffi and S. Isola {\em Classical
limit of the quantized hyperbolic toral automorphisms}, Comm. Math
Phys. {\bf 167} (1995), 471--507. 

\bibitem[HB]{HB} J.H. Hannay and M.V. Berry  {\em Quantization of linear
maps on a torus - Fresnel diffraction by a periodic grating}, 
Physica~D {\bf 1} (1980), 267--291. 


\bibitem[HR]{Hejhal}
D.A. Hejhal and B. Rackner {\em On the topography of Maass waveforms
for ${\rm PSL}(2,{Z})$}, Experiment. Math. {\bf 1} (1992), no. 4, 275--305. 

\bibitem[Ho]{Ho}
L. Hormander {\em The spectral function for an elliptic operator}, 
Acta Math. {\bf 127} (1968), pp. 193--218.

 \bibitem[IS]{IS}
 H. Iwaniec and P. Sarnak {\em $L\sp \infty$ norms of eigenfunctions of
 arithmetic surfaces}, Ann. of Math. (2) {\bf 141} (1995), no. 2, 301--320.

\bibitem[Ka]{Katz-recent}
N.M. Katz {\em Sato-Tate equidistribution of Kurlberg-Rudnick sums},
preprint, December 2000. 

 \bibitem[KR]{KR1}
 P. Kurlberg and Z. Rudnick {\em Hecke theory and equidistribution 
 for the quantization of linear maps of the torus}, 
 Duke Math. J. {\bf 103} (2000), no. 1, 47--77.



\bibitem[AQC]{Sarnak-AQC}
P. Sarnak {\em Arithmetic quantum chaos}. The Schur lectures (1992) (Tel Aviv),
183--236, Israel Math. Conf. Proc., 8, Bar-Ilan Univ., Ramat Gan, 1995. 

 \bibitem[S]{Sarnak-CM}
 P. Sarnak {\em Estimates for Rankin-Selberg $L$-functions and Quantum
 Unique Ergodicity}, preprint July 2000. 

\bibitem[Sch]{Schmidt-book} W. M. Schmidt {\em Equations over finite
fields: An elementary approach}, Lecture Notes in Math {\bf 536}
(1976) Springer-Verlag, Berlin Heidelberg. 




 \bibitem[W]{Watson}
 T. Watson, Princeton Ph.D. thesis, in progress. 

\bibitem[Weil]{Weil} 
A. Weil {\em Sur les Courbes Alg\'ebriques et les 
Vari\'et\'es qui s'en D\'eduisent}, Hermann, Paris (1948). 


 \end{thebibliography}
 \end{document}